\documentclass[prl,twocolumn,showpacs,superscriptaddress,floatfix]{revtex4}
 \def\LineNumbers{false}	
 \def\AuthorList{true}

\usepackage{ifthen}
\usepackage{lineno}
\usepackage{slashed}
\usepackage{epsfig}
\usepackage{amsmath}
\usepackage{comment}
\usepackage{longtable}
\usepackage{subfigure}
\usepackage{color}

\def\met{{\mbox{$E\kern-0.57em\raise0.19ex\hbox{/}_{T}$}}}

\def\bb{$b\bar{b}$}
\def\metbb{$WH,ZH\rightarrow\met b\bar{b}$}
\def\WH{$WH\rightarrow \ell\nu b\bar{b}$}

\def\lmet{$WH\rightarrow \ell\kern-0.45em\raise0.19ex\hbox{/} \nu b\bar{b}$}
\def\ZH{$ZH\rightarrow \ell^+\ell^- b\bar{b}$}

\newcommand{\METVEC}{\mbox{$\raisebox{.3ex}{$\not\!$}{\vec E}_T$}}

\newcommand{\MCFM}       {{\sc mcfm}}
\newcommand{\PYTHIA}     {{\sc pythia}}
\newcommand{\ALPGEN}       {{\sc alpgen}}

\newcommand{\HDECAY}       {{\sc hdecay}}

\newcommand{\rar}       {\rightarrow}
\newcommand{\MET}{$\not\!\!E_T$}

\begin{document}

\ifthenelse{\equal{\LineNumbers}{true}}{
 \pagewiselinenumbers
 \linenumbers
}

\title{Combined search for the standard model Higgs boson decaying to a $b\bar{b}$ pair using the full CDF data set}

\ifthenelse{\equal{\AuthorList}{true}}{
  \affiliation{Institute of Physics, Academia Sinica, Taipei, Taiwan 11529, Republic of China}
\affiliation{Argonne National Laboratory, Argonne, Illinois 60439, USA}
\affiliation{University of Athens, 157 71 Athens, Greece}
\affiliation{Institut de Fisica d'Altes Energies, ICREA, Universitat Autonoma de Barcelona, E-08193, Bellaterra (Barcelona), Spain}
\affiliation{Baylor University, Waco, Texas 76798, USA}
\affiliation{Istituto Nazionale di Fisica Nucleare Bologna, $^{ee}$University of Bologna, I-40127 Bologna, Italy}
\affiliation{University of California, Davis, Davis, California 95616, USA}
\affiliation{University of California, Los Angeles, Los Angeles, California 90024, USA}
\affiliation{Instituto de Fisica de Cantabria, CSIC-University of Cantabria, 39005 Santander, Spain}
\affiliation{Carnegie Mellon University, Pittsburgh, Pennsylvania 15213, USA}
\affiliation{Enrico Fermi Institute, University of Chicago, Chicago, Illinois 60637, USA}
\affiliation{Comenius University, 842 48 Bratislava, Slovakia; Institute of Experimental Physics, 040 01 Kosice, Slovakia}
\affiliation{Joint Institute for Nuclear Research, RU-141980 Dubna, Russia}
\affiliation{Duke University, Durham, North Carolina 27708, USA}
\affiliation{Fermi National Accelerator Laboratory, Batavia, Illinois 60510, USA}
\affiliation{University of Florida, Gainesville, Florida 32611, USA}
\affiliation{Laboratori Nazionali di Frascati, Istituto Nazionale di Fisica Nucleare, I-00044 Frascati, Italy}
\affiliation{University of Geneva, CH-1211 Geneva 4, Switzerland}
\affiliation{Glasgow University, Glasgow G12 8QQ, United Kingdom}
\affiliation{Harvard University, Cambridge, Massachusetts 02138, USA}
\affiliation{Division of High Energy Physics, Department of Physics, University of Helsinki and Helsinki Institute of Physics, FIN-00014, Helsinki, Finland}
\affiliation{University of Illinois, Urbana, Illinois 61801, USA}
\affiliation{The Johns Hopkins University, Baltimore, Maryland 21218, USA}
\affiliation{Institut f\"{u}r Experimentelle Kernphysik, Karlsruhe Institute of Technology, D-76131 Karlsruhe, Germany}
\affiliation{Center for High Energy Physics: Kyungpook National University, Daegu 702-701, Korea; Seoul National University, Seoul 151-742, Korea; Sungkyunkwan University, Suwon 440-746, Korea; Korea Institute of Science and Technology Information, Daejeon 305-806, Korea; Chonnam National University, Gwangju 500-757, Korea; Chonbuk National University, Jeonju 561-756, Korea}
\affiliation{Ernest Orlando Lawrence Berkeley National Laboratory, Berkeley, California 94720, USA}
\affiliation{University of Liverpool, Liverpool L69 7ZE, United Kingdom}
\affiliation{University College London, London WC1E 6BT, United Kingdom}
\affiliation{Centro de Investigaciones Energeticas Medioambientales y Tecnologicas, E-28040 Madrid, Spain}
\affiliation{Massachusetts Institute of Technology, Cambridge, Massachusetts 02139, USA}
\affiliation{Institute of Particle Physics: McGill University, Montr\'{e}al, Qu\'{e}bec, Canada H3A~2T8; Simon Fraser University, Burnaby, British Columbia, Canada V5A~1S6; University of Toronto, Toronto, Ontario, Canada M5S~1A7; and TRIUMF, Vancouver, British Columbia, Canada V6T~2A3}
\affiliation{University of Michigan, Ann Arbor, Michigan 48109, USA}
\affiliation{Michigan State University, East Lansing, Michigan 48824, USA}
\affiliation{Institution for Theoretical and Experimental Physics, ITEP, Moscow 117259, Russia}
\affiliation{University of New Mexico, Albuquerque, New Mexico 87131, USA}
\affiliation{The Ohio State University, Columbus, Ohio 43210, USA}
\affiliation{Okayama University, Okayama 700-8530, Japan}
\affiliation{Osaka City University, Osaka 588, Japan}
\affiliation{University of Oxford, Oxford OX1 3RH, United Kingdom}
\affiliation{Istituto Nazionale di Fisica Nucleare, Sezione di Padova-Trento, $^{ff}$University of Padova, I-35131 Padova, Italy}
\affiliation{University of Pennsylvania, Philadelphia, Pennsylvania 19104, USA}
\affiliation{Istituto Nazionale di Fisica Nucleare Pisa, $^{gg}$University of Pisa, $^{hh}$University of Siena and $^{ii}$Scuola Normale Superiore, I-56127 Pisa, Italy}
\affiliation{University of Pittsburgh, Pittsburgh, Pennsylvania 15260, USA}
\affiliation{Purdue University, West Lafayette, Indiana 47907, USA}
\affiliation{University of Rochester, Rochester, New York 14627, USA}
\affiliation{The Rockefeller University, New York, New York 10065, USA}
\affiliation{Istituto Nazionale di Fisica Nucleare, Sezione di Roma 1, $^{jj}$Sapienza Universit\`{a} di Roma, I-00185 Roma, Italy}
\affiliation{Rutgers University, Piscataway, New Jersey 08855, USA}
\affiliation{Texas A\&M University, College Station, Texas 77843, USA}
\affiliation{Istituto Nazionale di Fisica Nucleare Trieste/Udine, I-34100 Trieste, $^{kk}$University of Udine, I-33100 Udine, Italy}
\affiliation{University of Tsukuba, Tsukuba, Ibaraki 305, Japan}
\affiliation{Tufts University, Medford, Massachusetts 02155, USA}
\affiliation{University of Virginia, Charlottesville, Virginia 22906, USA}
\affiliation{Waseda University, Tokyo 169, Japan}
\affiliation{Wayne State University, Detroit, Michigan 48201, USA}
\affiliation{University of Wisconsin, Madison, Wisconsin 53706, USA}
\affiliation{Yale University, New Haven, Connecticut 06520, USA}

\author{T.~Aaltonen}
\affiliation{Division of High Energy Physics, Department of Physics, University of Helsinki and Helsinki Institute of Physics, FIN-00014, Helsinki, Finland}
\author{B.~\'{A}lvarez~Gonz\'{a}lez$^z$}
\affiliation{Instituto de Fisica de Cantabria, CSIC-University of Cantabria, 39005 Santander, Spain}
\author{S.~Amerio}
\affiliation{Istituto Nazionale di Fisica Nucleare, Sezione di Padova-Trento, $^{ff}$University of Padova, I-35131 Padova, Italy}
\author{D.~Amidei}
\affiliation{University of Michigan, Ann Arbor, Michigan 48109, USA}
\author{A.~Anastassov$^x$}
\affiliation{Fermi National Accelerator Laboratory, Batavia, Illinois 60510, USA}
\author{A.~Annovi}
\affiliation{Laboratori Nazionali di Frascati, Istituto Nazionale di Fisica Nucleare, I-00044 Frascati, Italy}
\author{J.~Antos}
\affiliation{Comenius University, 842 48 Bratislava, Slovakia; Institute of Experimental Physics, 040 01 Kosice, Slovakia}
\author{G.~Apollinari}
\affiliation{Fermi National Accelerator Laboratory, Batavia, Illinois 60510, USA}
\author{J.A.~Appel}
\affiliation{Fermi National Accelerator Laboratory, Batavia, Illinois 60510, USA}
\author{T.~Arisawa}
\affiliation{Waseda University, Tokyo 169, Japan}
% ...

\collaboration{CDF Collaboration\footnote[2]{With visitors from
$^a$Istituto Nazionale di Fisica Nucleare, Sezione di Cagliari, 09042 Monserrato (Cagliari), Italy,
$^b$University of CA Irvine, Irvine, CA 92697, USA,
$^c$University of CA Santa Barbara, Santa Barbara, CA 93106, USA,
$^d$University of CA Santa Cruz, Santa Cruz, CA 95064, USA,
$^e$Institute of Physics, Academy of Sciences of the Czech Republic, Czech Republic,
$^f$CERN, CH-1211 Geneva, Switzerland,
$^g$Cornell University, Ithaca, NY 14853, USA,
$^h$University of Cyprus, Nicosia CY-1678, Cyprus,
$^i$Office of Science, U.S. Department of Energy, Washington, DC 20585, USA,
$^j$University College Dublin, Dublin 4, Ireland,
$^k$ETH, 8092 Zurich, Switzerland,
$^l$University of Fukui, Fukui City, Fukui Prefecture, Japan 910-0017,
$^m$Universidad Iberoamericana, Mexico D.F., Mexico,
$^n$University of Iowa, Iowa City, IA 52242, USA,
$^o$Kinki University, Higashi-Osaka City, Japan 577-8502,
$^p$Kansas State University, Manhattan, KS 66506, USA,
$^q$Ewha Womans University, Seoul, 120-750, Korea,
$^r$University of Manchester, Manchester M13 9PL, United Kingdom,
$^s$Queen Mary, University of London, London, E1 4NS, United Kingdom,
$^t$University of Melbourne, Victoria 3010, Australia,
$^u$Muons, Inc., Batavia, IL 60510, USA,
$^v$Nagasaki Institute of Applied Science, Nagasaki, Japan,
$^w$National Research Nuclear University, Moscow, Russia,
$^x$Northwestern University, Evanston, IL 60208, USA,
$^y$University of Notre Dame, Notre Dame, IN 46556, USA,
$^z$Universidad de Oviedo, E-33007 Oviedo, Spain,
$^{aa}$CNRS-IN2P3, Paris, F-75205 France,
$^{bb}$Texas Tech University, Lubbock, TX 79609, USA,
$^{cc}$Universidad Tecnica Federico Santa Maria, 110v Valparaiso, Chile,
$^{dd}$Yarmouk University, Irbid 211-63, Jordan.
}}
\noaffiliation
 
}

\ifthenelse{\equal{\AuthorList}{false}}{
\begin{figure}  
  \leftline{\includegraphics[scale=0.5]{cdfii_thumb_logo.eps}\hfill
    CDF/PUB/EXOTIC/CDFR/10880}
\end{figure}

\vspace*{0.5cm} 

\author{The CDF Collaboration}
\affiliation{URL http://www-cdf.fnal.gov}
}

\date{\today}

\begin{abstract}

\ifthenelse{\equal{\AuthorList}{false}}{
\vspace*{0.2cm}
}

We combine the results of searches for the standard model Higgs boson based on the full CDF Run II data set obtained from $\sqrt{s}=1.96$~TeV $p{\bar{p}}$ collisions at the Fermilab Tevatron corresponding to an integrated luminosity of $9.45~\mathrm{fb^{-1}}$. The searches are conducted for Higgs bosons that are produced in association with a $W$ or $Z$ boson, have masses in the range 90--150~GeV$/c^2$, and decay into $b{\bar{b}}$ pairs.  An excess of data is present that is inconsistent with the background prediction at the level of 2.5 standard deviations (the most significant local excess is 2.7 standard deviations).

\end{abstract}

\pacs{13.85.Rm, 14.80.Bn}

\maketitle

The mechanism of electroweak symmetry breaking in the standard model (SM)~\cite{sm, higgs} predicts the existence of a fundamental scalar boson, referred to as the Higgs boson ($H$).  Although there is strong evidence of electroweak symmetry breaking, the Higgs boson has yet to be observed.  The SM does not predict the mass of the Higgs boson, $m_H$, but the combination of precision electroweak measurements~\cite{elweak}, including recent top quark and $W$ boson mass measurements from the Tevatron~\cite{topmass,wmass}, constrains $m_H<152$~GeV/$c^2$ at the 95\% confidence level.  Direct searches at LEP2~\cite{sm-lep}, the Tevatron~\cite{tevcomb2012}, and the LHC~\cite{cmscomb2012,atlascomb2012} exclude all possible masses of the SM Higgs boson at the 95\% confidence level or the 95\% credibility level (C.L.), except within the ranges 116.6 -- 119.4~GeV/$c^2$ and 122.1 -- 127~GeV/$c^2$.  A SM Higgs boson in these mass ranges would be produced in the $\sqrt{s}=1.96$ TeV $p{\bar{p}}$ collisions of the Tevatron and have a branching fraction to $b \bar{b}$ greater than 50\%~\cite{vhtheory, stange, lhcdifferential}.  While the most sensitive searches for the SM Higgs boson at the LHC are those based on its decays into pairs of gauge bosons, searches based on decays into pairs of $b$ quarks are the most sensitive at the Tevatron.  The searches at the LHC in the four-lepton and diphoton final state offer precise measurements of the mass of the Higgs boson, while the results presented here provide direct information about the Higgs boson's couplings to $b$ quarks and are therefore complementary to the primary LHC search modes.  In searches for the production of a Higgs boson in association with a vector boson ({\it WH } or {\it ZH }), leptonic decays of the vector boson provide effective discrimination between the expected signal and the large, uncertain hadronic backgrounds. Previous Higgs searches focused on these production and decay modes have been performed at LEP2~\cite{sm-lep} and the LHC~\cite{cmsbb,atlasbb}.  This Letter describes the combination of the results of three CDF searches for a SM-like Higgs boson with a mass in the range $90<m_H<150$~GeV/$c^2$.  These searches are targeted at \ZH{} ~\cite{llbb}, \WH{} ~\cite{lvbb}, and \metbb{} ~\cite{metbb}.

The CDF~II detector is described in detail elsewhere~\cite{cdf, coord}.  Calorimeter energy deposits are clustered into jets using a cone algorithm with an opening angle of $\Delta R=\sqrt{(\Delta\phi)^2+(\Delta\eta)^2}=0.4$~\cite{Bhatti:2005ai}. High-$p_T$ electron candidates are identified by matching charged-particle tracks in the tracking systems~\cite{silicon, cot} with energy deposits in the electromagnetic calorimeters~\cite{em}.  Muon candidates are identified by matching tracks with muon-detector track segments~\cite{muons}.  The hermeticity of the calorimeter allows for good reconstruction of the missing transverse energy (\met)~\cite{had}. Jets are identified as consistent with the fragmentation of a $b$ quark ($b$-tagged) using three different algorithms described in Ref.~\cite{tagging}, which make use of track impact parameters, the presence of identified displaced vertices, the presence of leptons near the jet, and jet kinematic properties.  The average tag efficiency for a jet originating (not originating) from $b$ quark fragmentation is in the range 42--70\% (0.9--8.9\%), depending on the properties of the jet.

Higgs boson signal events are simulated using \PYTHIA~\cite{pythia}, with CTEQ5L~\cite{cteq5l} parton distribution functions (PDFs) at leading order (LO). We normalize our Higgs boson signal-rate predictions to the most recent higher-order calculations available. The {\it WH} and {\it ZH} cross section calculations are performed at next-to-next-to leading order (NNLO) precision in QCD and next-to-leading-order (NLO) precision in the electroweak corrections and are described in Ref.~\cite{vhtheory}. The branching fractions for the Higgs boson decays are obtained from Ref.~\cite{lhcdifferential}.  These rely on calculations using \HDECAY~\cite{hdecay} and {\sc prophecy4f}~\cite{prophecy4f}.  Assuming the $m_H$ = 125~GeV/$c^2$ hypothesis, we expect approximately 85 Higgs boson events to pass our selections.  We model SM and instrumental background processes using a mixture of Monte Carlo (MC) and data-driven methods.  Diboson ({\it WW }, {\it WZ }, {\it ZZ }) MC samples are normalized using the NLO calculations from \MCFM~\cite{mcfm}.  For $t{\bar{t}}$ we use a production cross section of $7.04\pm 0.7$~pb~\cite{mochuwer}, which is based on a top-quark mass of 173~GeV/$c^2$~\cite{topmass} and MSTW 2008 NNLO PDFs~\cite{MSTW}.  The single-top-quark production cross section is taken to be $3.15\pm 0.31$~pb~\cite{kidonakis_st}. The normalization of the $Z$+jets and $W$+jets MC samples is taken from \ALPGEN{}~\cite{Mangano:2002ea} corrected for NLO effects, except in the case of the \WH{} search.  The normalization of the $W$+jets MC sample in the \WH{} search, and normalization of the instrumental and QCD multi-jet samples in all searches, are constrained from data samples where the expected signal is several orders of magnitude smaller than in the search samples.

All searches use the same data sample, which corresponds to 9.45 fb$^{-1}$ of integrated luminosity~\cite{lumi_grl}. The analysis channels select non-overlapping subsets of the data.  Exactly two, one, or zero charged leptons are required by the \ZH{}, \WH{}, and \metbb{} event selections, respectively, where $\ell$ denotes a reconstructed electron or muon.  Both the \WH{} and \metbb{} event selections require large \met{} to be consistent with the signature of one or more high-$p_T$ neutrinos escaping the detector.  Events in all searches are required to contain exactly two or three reconstructed jets.  To optimize the sensitivity, the data in each search are further divided into independent sub-channels composed of differing jet multiplicity, lepton quality, $b$-tag multiplicity, and $b$-tag quality.  There are 16, 26, and 3 sub-channels for the \ZH{}, \WH{}, and \metbb{} analyses, respectively, totaling to 45 for the combination presented here. For a pair of jets, the dijet mass resolution for signal events at CDF is expected to be 10--15\% of their mean reconstructed mass~\cite{Aaltonen:2011di}.  The decay width of the Higgs boson signal is predicted to be much smaller than this mass resolution. The presence of a signal would appear as a broad enhancement in the invariant mass distribution of jets.  The dijet mass provides the greatest discrimination between signal and background.  However, to enhance sensitivity the dijet mass is combined with other kinematic information into multivariate discriminants.

  Each search sub-channel uses a multivariate analysis (MVA) technique
designed to separate the Higgs boson signal from the backgrounds.  The
MVA functions are optimized separately for each subchannel and for 13
independent mass hypotheses at each value of $m_H$ in the range
90 -- 150~GeV/$c^2$, in 5~GeV/$c^2$ intervals.  We interpret the
results using a Bayesian technique, separately at each value of $m_H$,
using a combined likelihood formed from a product of likelihoods for
the individual channels, each of which is a product over histogram
bins of the MVA outputs,
\begin{equation}
{\cal{L}}(R,{\vec{s}},{\vec{b}}|{\vec{n}},{\vec{\theta}})\times\pi({\vec{\theta}})
= \prod_{i=1}^{N_{\rm C}}\prod_{j=1}^{N_{\rm bins}} \mu_{ij}^{n_{ij}} \frac{e^{-\mu_{ij}}}{n_{ij}!}
\times\prod_{k=1}^{n_{\rm sys}}e^{-\theta_k^2/2}.
\end{equation}
\noindent In this expression, the first product is over the number of channels
($N_{\rm C}$), and the second product is over histogram bins containing $n_{ij}$ events, binned in  ranges of the final discriminant variables used for the individual analyses.   The predictions for the bin contents are $\mu_{ij} =R \times s_{ij}({\vec{\theta}}) + b_{ij}({\vec{\theta}})$ for channel $i$ and histogram bin $j$, where $s_{ij}$ and $b_{ij}$ represent the expected SM signal and background in the bin, and $R$ is a scaling factor applied to the signal.  By scaling all signal contributions by the same factor, we assume that the relative contributions of the different processes are as given by the SM.

Systematic uncertainties are parametrized by the dependence of $s_{ij}$ and $b_{ij}$ on ${\vec\theta}$.  Each of the $n_{\rm sys}$ components of ${\vec\theta}$, $\theta_k$, corresponds to a single independent source of systematic uncertainty, and each parameter may have an impact on several sources of signal and background in different channels, thus accounting for correlations.  Gaussian priors are assumed for the $\theta_k$, truncated so that no prediction is
negative.  The likelihood function, multiplied by the $\theta_k$ priors, $\pi(\theta_k)$, is then  integrated over $\theta_k$ including correlations~\cite{pdgstats},
\begin{equation}
{\cal{L}}^\prime(R) = \int{\cal{L}}(R,{\vec{s}},{\vec{b}}|{\vec{n}},{\vec{\theta}})\pi({\vec{\theta}})d{\vec{\theta}}.
\end{equation}

We assume a uniform prior in $R$ to obtain its posterior distribution.
The observed 95\% C.L. upper limit on $R$, $R_{95}^{\rm{obs}}$,
satisfies $0.95=\int_0^{R_{95}^{\rm{obs}}}\cal{L}^\prime(R) \mbox{d}R$.  The
expected distribution of $R_{95}$ is computed in an ensemble of
pseudoexperiments generated without signal.  In each pseudoexperiment,
random values of the nuisance parameters are drawn from their priors.
The median expected value of $R_{95}$ in this ensemble is denoted
$R_{95}^{\rm{exp}}$.  A combined measurement of the cross section for
Higgs boson production assuming SM branching ratios in units of the SM
production rates is given by $R^{\rm{fit}}$, which is the value of $R$
that maximizes ${\cal L}^\prime$. The 68\% C.L. interval (one standard deviation) is quoted as
the shortest interval containing 68\% of the integral of the
posterior.

Though many sources of systematic uncertainty differ among the analyses, all correlations are taken into account in the combined limits, cross sections, and $p$-values. The uncertainties on the signal production cross sections are estimated from the factorization and renormalization scale variations, which includes the impact of uncalculated higher-order corrections, uncertainties due to PDFs, and the dependence on the strong coupling constant, ($\alpha_s$). The resulting uncertainties on the inclusive {\it WH} and {\it ZH} production rates are 5\%~\cite{vhtheory}. We assign uncertainties to the Higgs boson decay branching ratios as calculated in Ref.~\cite{denner11}.  These uncertainties arise from imperfect knowledge of the mass of the $b$ and $c$ quarks, $\alpha_s$, and theoretical uncertainties in the $b{\bar{b}}$ decay rates. The largest sources of uncertainty on the dominant backgrounds in the $b$-tagged channels are the rates of $V$+heavy flavor jets, where $V=W$~or~$Z$, which are typically 30\% of the predicted values. The posterior uncertainties on these rates are typically 8\% or less.  Because the different analyses use different methods to obtain the $V$+heavy flavor predictions, we treat their uncertainties as uncorrelated between the $\ell\nu b{\bar{b}}$, the $\met b{\bar{b}}$, and $\ell^+\ell^-b{\bar{b}}$ channels. We use simulated events to study the impact of the jet energy scale uncertainty~\cite{Bhatti:2005ai} on the rates and shapes of the signal and background expectations. We observe that the jet energy scale uncertainty is highly constrained by the data in the individual channels.  Because differences between channels in the event selection and modeling of the background shapes affect the constraint on the jet energy scale obtained from the fit, we conservatively choose to treat the jet energy scale variations uncorrelated between the three analyses in the combined search.

Uncertainties on lepton identification and trigger efficiencies range from 2\% to 6\% and are applied to both signal- and MC-based background predictions. The uncertainty on the integrated luminosity of 6\% arises from uncertainties in the luminosity monitor acceptance and the inelastic $p{\bar{p}}$ cross section~\cite{inelppbar}, and is assumed to be correlated between the signal- and MC-based background predictions.

 \begin{figure}[htb]
 \begin{centering}
 \includegraphics[width=0.8\columnwidth]{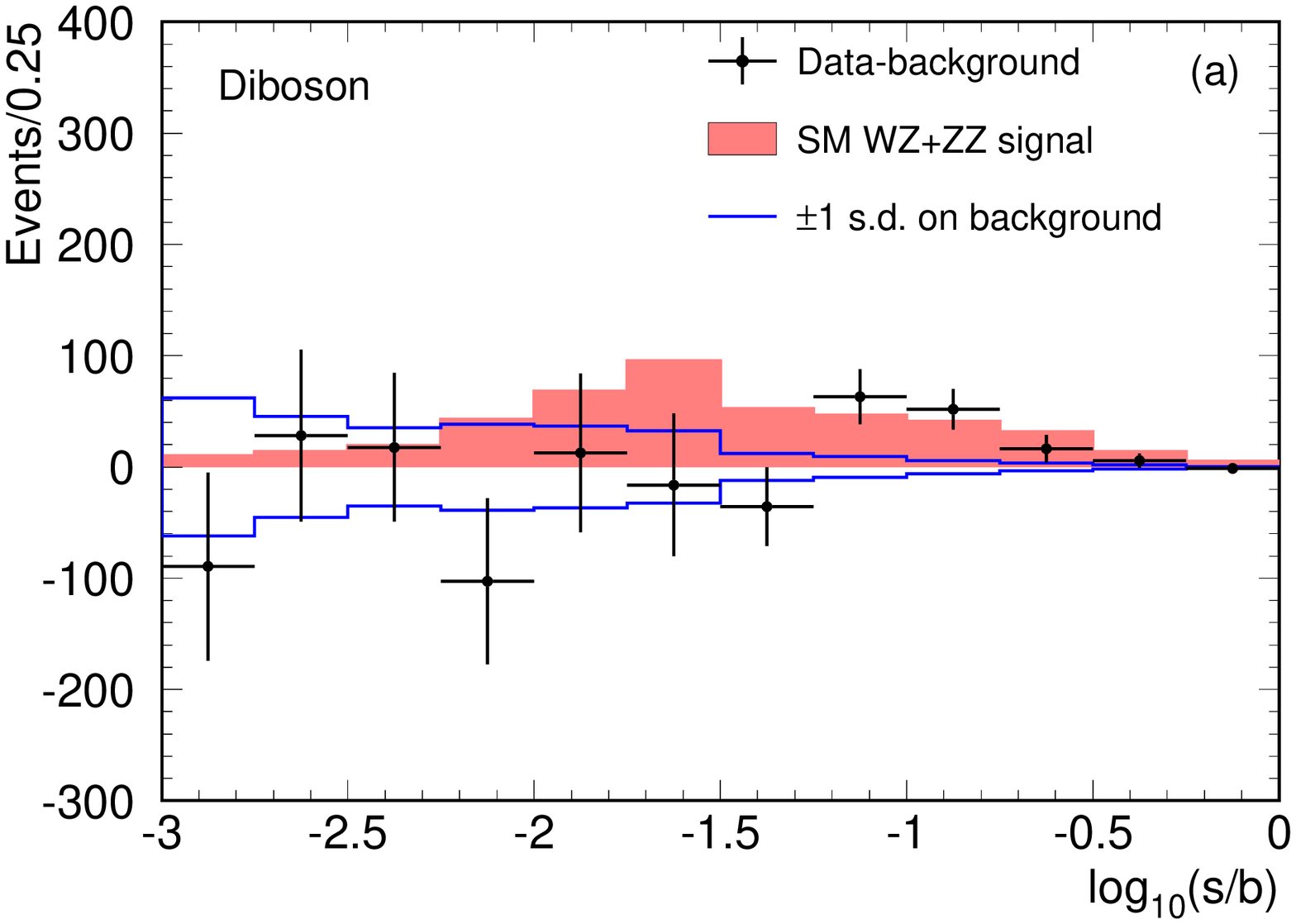}
 \includegraphics[width=0.8\columnwidth]{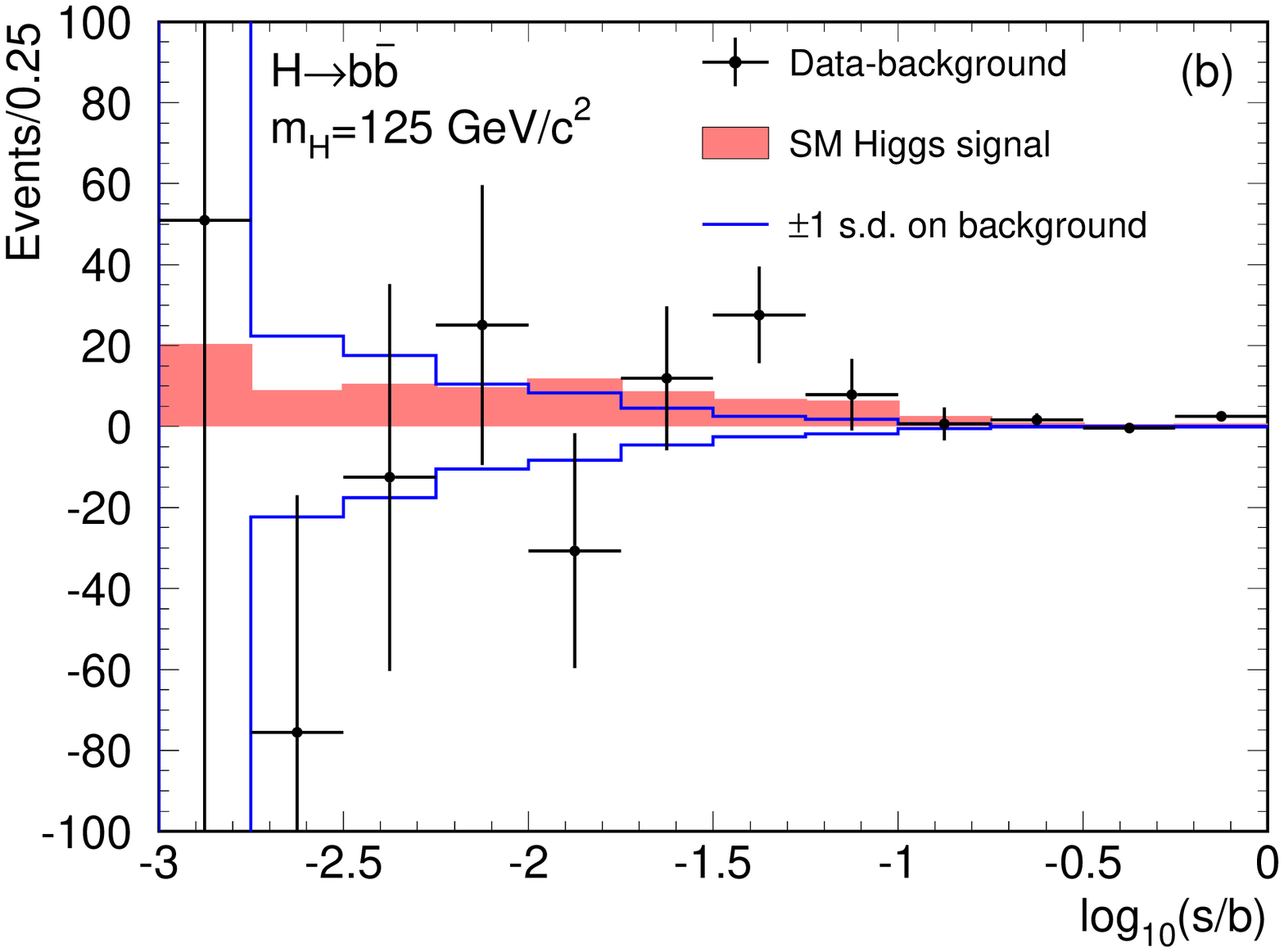}
 \caption{
 \label{fig:bgsub}
Background-subtracted distributions for the discriminant histograms, summed for bins
with similar signal-to-background ratio ($s/b$), for the diboson search (top) and the $H\rightarrow b{\bar{b}}$ $(m_H=125$~GeV/$c^2)$ search (bottom). The background has been fit to the data, and the uncertainty on the background
is the post-fit uncertainty.  The signal model, which is normalized to the SM expectation,
is shown with a filled histogram.  The uncertainty on the background-subtracted data points shown is the square root of the post-fit background prediction in each bin.  The leftmost bin contains all events of lower $s/b$ values.
}
 \end{centering}
 \end{figure}

\begin{figure}[t]
\begin{centering}
\includegraphics[width=7.4cm]{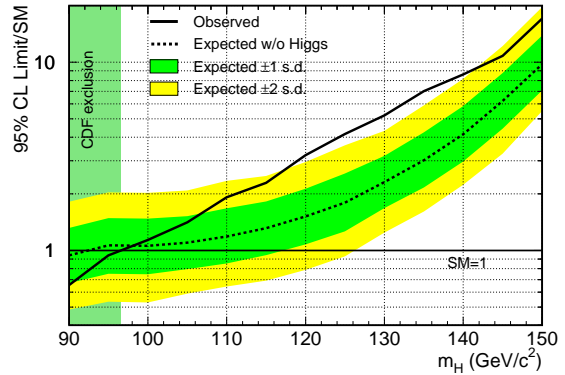}
\caption{
\label{fig:limits}
Observed and expected 95\% C.L. upper limits on SM Higgs boson production (R$_{95}$) as a function of Higgs boson mass.
The shaded bands indicate the credibility bands in which $R_{95}$ is expected to
fluctuate, in the absence of signal.
}
\end{centering}
\end{figure}

 \begin{figure}[htb]
 \begin{centering}
 \includegraphics[width=7.4cm]{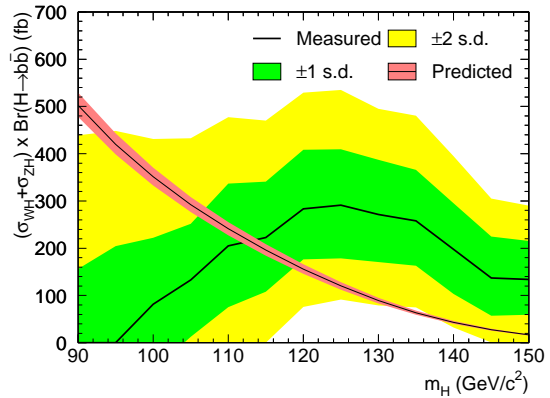}
 \caption{
 \label{fig:xsfit}
The best-fit cross section times branching ratio $(\sigma_{WH}+\sigma_{ZH})\times \mathcal{B}(H\rightarrow b{\bar{b}})$
as a function of $m_H$.  The dark-shaded region shows the one standard deviation C.L. band, the light-shaded region shows the
two standard deviation C.L. region, and the SM prediction is shown as the smooth, falling curve with a narrow band indicating the theoretical
uncertainty. }
 \end{centering}
 \end{figure}

 \begin{figure}[htb]
 \begin{centering}
 \includegraphics[width=7.4cm]{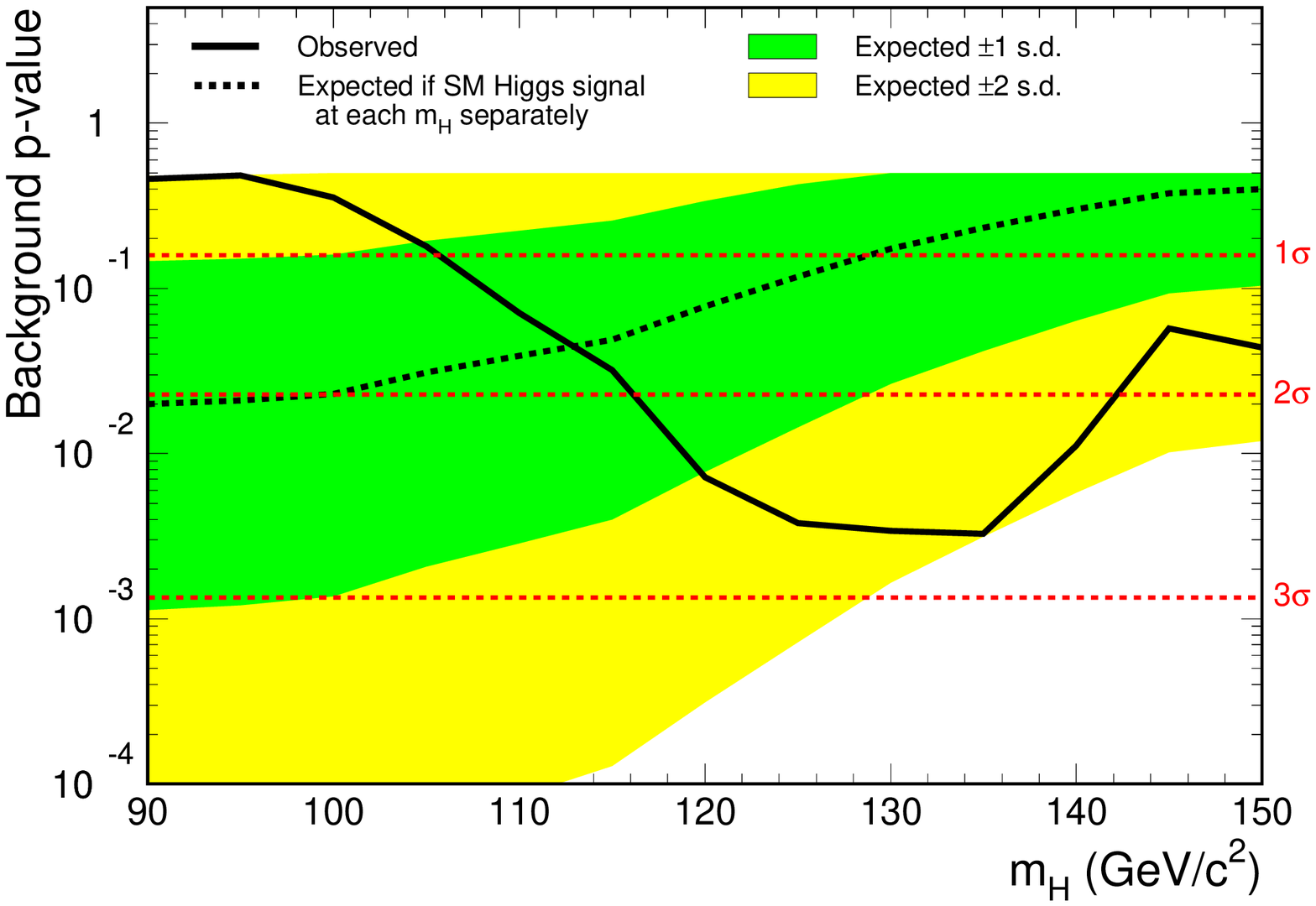}
 \caption{
 \label{fig:pvalue}
Background-only $p$-value for the combined search.  Also shown are the median expected values
and the one and two standard deviation C.L. bands assuming a SM signal is present, evaluated at each $m_H$ separately.
}
 \end{centering}
 \end{figure}

To validate our background modeling and search methods, we additionally perform a search for SM diboson production in the same final states used for the SM $H\rar b\bar{b}$ searches.   The NLO SM cross section for {\it VZ} times the branching fraction of $Z\rar b\bar{b}$ is 682 $\pm$ 50~fb, which is comparable to the 410 $\pm$ 20~fb cross section times branching fraction of $VH(H\rar b\bar{b})$ for a 100~GeV/$c^2$ SM Higgs boson.  The data sample, reconstruction, background models, uncertainties, and subchannel divisions are identical to those of the SM Higgs boson search, but the discriminant functions are trained specifically for the signal of SM diboson production.  The measured cross section for $VZ$ is $4.1 \pm 1.3$~pb (stat+syst), which is consistent with the SM prediction of $4.4 \pm 0.3$~pb and corresponds to a diboson signal significance of $\sim$3.2 standard deviations.

To better visualize the data, we combine the histograms of the final
discriminants, adding the contents of bins with similar
signal-to-background ratio ($s/b$).  Figure~\ref{fig:bgsub} shows the
signal expectation and the data with the background subtracted, as a
function of the $s/b$ of the collected bins, for the diboson analysis
described above and for the combined Higgs boson search, assuming $m_H=125$~GeV/$c^2$.  The
background model has been fit to the data, and the uncertainties on
the background are those after the nuisance parameters have been
constrained in the fit.  An excess of Higgs boson candidate events in the highest $s/b$
bins relative to the background-only expectation is observed in Fig.~\ref{fig:bgsub}.

We extract limits on SM Higgs boson production in the $m_H$ range of 90--150~GeV/$c^2$.  We
present our results in terms of $R_{95}^{\mathrm{obs}}$, the ratio of
the limits obtained to the rate predicted by the SM, as a function of
the Higgs boson mass.  We assume the SM ratio for {\it WH} and {\it
ZH} production.  A value of $R_{95}^{\mathrm{obs}}$ less than or equal
to one indicates a SM Higgs boson mass that is excluded at the 95\%
C.L.  These limits are shown, together with the median expected values
and distributions of individual experiments assuming a signal is absent in
Fig.~\ref{fig:limits}.  We also compute the best-fit rate parameter
$R^{\rm{fit}}$, which, when multiplied by the SM prediction for the
associated production cross section times the decay branching ratio
($\sigma_{WH}+\sigma_{ZH})\times \mathcal{B}(H\rightarrow
b{\bar{b}})$, yields the best fit values for this product. We show our
fitted ($\sigma_{WH}+\sigma_{ZH})\times \mathcal{B}(H\rightarrow
b{\bar{b}})$ as a function of $m_H$, along with the SM prediction, in
Fig.~\ref{fig:xsfit}.

Significances of excesses in data over the background prediction are
computed by calculating the local background-only $p$-value using
$R^{\rm{fit}}$ as the test statistic.  This $p$-value is the
probability that $R^{\rm{fit}}$ is equal to or exceeds its observed
value, assuming a signal is truly absent.  The Look-Elsewhere Effect
(LEE)~\cite{LEE,Dunn1961Multiple} accounts for the possibility of a background
fluctuation affecting the local $p$-value anywhere in the tested $m_H$
range, here taken to be from 115 to 150~GeV/$c^2$, owing to the prior
exclusion~\cite{sm-lep}.  In this mass range, the reconstructed mass
resolution is approximately 15-20 GeV/$c^2$. We therefore estimate
that two independent outcomes are possible in these searches
(LEE factor~$\approx$~2).  The $p$-value is computed for each $m_H$ in the range
90--150~GeV/$c^2$, and is shown in Fig.~\ref{fig:pvalue}.  Also shown
are the expected values of the $p$-value assuming a SM signal is
present, testing each value of $m_H$ in turn.  The maximum local
significance corresponds to 2.7 standard deviations at
$m_H=135$~GeV/$c^2$. Correcting for the LEE yields a global
significance of 2.5 standard deviations.

In summary, we present a combination of CDF searches for the SM Higgs boson decaying to \bb{} pairs using the entire Run~II data sample.  We search for a Higgs boson with a mass between 90 and 150~GeV/$c^2$, and exclude Higgs bosons with masses smaller than 96~GeV/$c^2$.  The observed credibility limits are higher than those expected in the background-only hypothesis in the mass range 115--150~GeV/$c^2$.  Within the currently non-excluded mass ranges, the lowest local $p$-value is found for a Higgs boson mass of 125~GeV/$c^2$, where the local significance of this deviation with respect to the background-only hypothesis is 2.7 standard deviations.  At the same mass hypothesis, we measure an associated production cross section times the decay branching ratio of $(\sigma_{WH}+\sigma_{ZH})\times \mathcal{B}(H\rightarrow b{\bar{b}})=  291 ^{+118}_{-113} ~(\mathrm{stat+sys})$~fb.

   This result is of fundamental interest both because similar searches are difficult at the LHC and because verification of a Higgs-boson-like particle decaying to $b$ quarks would offer a measurement of the $b$-quark Yukawa coupling, further establishing the mechanism of electroweak symmetry breaking as the source of fermionic mass in the quark sector.


We thank the Fermilab staff and the technical staffs of the
participating institutions for their vital contributions. This work
was supported by the U.S. Department of Energy and National Science
Foundation; the Italian Istituto Nazionale di Fisica Nucleare; the
Ministry of Education, Culture, Sports, Science and Technology of
Japan; the Natural Sciences and Engineering Research Council of
Canada; the National Science Council of the Republic of China; the
Swiss National Science Foundation; the A.P. Sloan Foundation; the
Bundesministerium f\"ur Bildung und Forschung, Germany; the Korean
World Class University Program, the National Research Foundation of
Korea; the Science and Technology Facilities Council and the Royal
Society, UK; the Russian Foundation for Basic Research; the Ministerio
de Ciencia e Innovaci\'{o}n, and Programa Consolider-Ingenio 2010,
Spain; the Slovak R\&D Agency; the Academy of Finland; and the
Australian Research Council (ARC).

\bibliographystyle{apsrev}


\begin{thebibliography}{42}
\bibitem[{coo()}]{coord}
\bibinfo{note}{We use a cylindrical coordinate system with its origin in the
  center of the detector, where $\theta$ and $\phi$ are the polar and azimuthal
  angles, respectively, and pseudorapidity is $\eta= - \rm ln$ $\rm
  tan(\theta/2)$. The missing $E_T$ (\METVEC) is defined by \METVEC$ =
  -\sum_iE_T^i{\hat n}_i, i={\rm calorimeter~tower~number}$, where ${\hat n}_i$
  is a unit vector perpendicular to the beam axis and pointing at the $ith$
  calorimeter tower. ~\METVEC{} is corrected for high-energy muons and also jet
  energy corrections. We define \MET=$|$\METVEC$|$. The transverse momentum
  $p_T$ is defined to be $p\sin\theta$.}
\bibitem[{lum()}]{lumi_grl}
\bibinfo{note}{The luminosity quoted here is that for data collected after
  reconstruction-specific detector requirements specific to each subchannel are
  applied.}
\end{thebibliography}

\end{document}